\documentclass[english,notitlepage,twocolumn,showpacs,prl]{revtex4-1}
\usepackage[T1]{fontenc}
\usepackage[latin9]{inputenc}
\usepackage{amsmath}
\usepackage{amssymb}
\usepackage{esint}
\usepackage{babel}
\begin{document}

\title{Contact theory for spin-orbit-coupled Fermi gases}

\author{Shi-Guo Peng$^{1}$}

\author{Cai-Xia Zhang$^{1,4}$}

\author{Shina Tan$^{2,3}$}
\email{shina.tan@physics.gatech.edu}

\author{Kaijun Jiang$^{1,3}$}
\email{kjjiang@wipm.ac.cn}

\affiliation{$^{1}$State Key Laboratory of Magnetic Resonance and Atomic and
Molecular Physics, Wuhan Institute of Physics and Mathematics, Chinese
Academy of Sciences, Wuhan 430071, China}

\affiliation{$^{2}$School of Physics, Georgia Institute of Technology, Atlanta,
Georgia 30332, USA}

\affiliation{$^{3}$Center for Cold Atom Physics, Chinese Academy of Sciences,
Wuhan 430071, China}

\affiliation{$^{4}$School of Physics, University of Chinese Academy of Sciences,
Beijing 100049, China}

\date{\today}
\begin{abstract}
We develop the contact theory for spin-orbit-coupled Fermi gases.
By using a perturbation method, we derive analytically the universal
two-body behavior at short distance, which does not depend on the
short-range details of interatomic potentials. We find that two new
scattering parameters need to be introduced because of spin-orbit
coupling, besides the traditional $s$- and $p$-wave scattering length
(volume) and effective ranges. This is a general and unique feature
for spin-orbit-coupled systems. Consequently, two new adiabatic energy
relations with respect to the new scattering parameters are obtained,
in which a new contact is involved because of spin-orbit coupling.
In addition, we derive the asymptotic behavior of the large-momentum
distribution, and find that the subleading tail is corrected by the
new contact. This work paves the way for exploring the profound properties
of spin-orbit-coupled many-body systems, according to two-body solutions. 
\end{abstract}
\maketitle
\emph{Introduction}.\textemdash Universality, referring to observations
independent of short-range details, is one of the most fascinating
and intriguing phenomena in modern physics. In ultracold atoms, a
set of universal relations, following from the short-range behavior
of the two-body physics are discovered \cite{Tan2008}. These relations
are connected simply by a universal contact parameter, which overarches
between microscopic and macroscopic properties of a strongly interacting
many-body system. Nowadays, the contact theory becomes significantly
important in ultracold atomic physics, and has systematically been
verified and investigated both experimentally and theoretically \cite{Zwerger2011,Yu2015U,He2016C,Luciuk2016E,Peng2016L,Fletcher2017Q}.
Nevertheless, the contact theory for spin-orbit-coupled systems is
still unexplored till now, even though the spin-orbit (SO) coupling
was realized in cold atoms several years ago \cite{Lin2011S,Wang2012S,Cheuk2012S},
and resulted unique phenomena have attracted a great deal of interest,
such as topological insulators and superconductors \cite{Qi2010T,Hasan2010T,Dalibard2011A,Zhang2014F}.

In this letter, for the first time, we generalize the contact theory
to strongly interacting spin-orbit-coupled Fermi gases, and the single-particle
Hamiltonian takes the form,
\begin{equation}
\hat{\mathcal{H}}_{1}=\frac{\hbar^{2}\hat{{\bf k}}_{1}^{2}}{2M}+\frac{\hbar^{2}\lambda}{M}\hat{{\bf k}}_{1}\cdot\hat{\boldsymbol{\sigma}}+\frac{\hbar^{2}\lambda^{2}}{2M},\label{eq:intr1}
\end{equation}
 where $\hat{{\bf k}}_{1}=-i\nabla$ and $\hat{\boldsymbol{\sigma}}$
are respectively the single-particle momentum and spin operators,
$\lambda>0$ is the strength of SO coupling, $M$ is the atomic mass,
and $\hbar$ is the Planck's constant divided by $2\pi$. Here, the
SO coupling is assumed to be isotropic for simplicity, and the possible
scheme for the realization of the three-dimensional (3D) isotropic
SO coupling is proposed in \cite{Anderson2012S}. Because of SO coupling,
the orbital angular momentum of the relative motion of two fermions
is no longer conserved, and then all the partial-wave scatterings
are coupled \cite{Cui2012M}. Fortunately, the total momentum ${\bf K}$
of two fermions is still conserved as well as the total angular momentum
${\bf J}$. Therefore, we may reasonably focus on the two-body problem
in the subspace of ${\bf K}=0$ and ${\bf J}=0$ for simplicity, and
then only $s$- and $p$-wave scatterings are coupled \cite{Cui2012M,Wu2013S}.
Consequently, the two spin-half fermions in the subspace of ${\bf K}=0$
and ${\bf J}=0$ is described by the following two-body Hamiltonian
\begin{equation}
\hat{\mathcal{H}}_{2}=\frac{\hbar^{2}\hat{{\bf k}}^{2}}{M}+\frac{\hbar^{2}\lambda}{M}\hat{{\bf k}}\cdot\left(\hat{\boldsymbol{\sigma}}_{2}-\hat{\boldsymbol{\sigma}}_{1}\right)+\frac{\hbar^{2}\lambda^{2}}{M}+V\left(r\right),\label{eq:2bHam}
\end{equation}
where $\hat{{\bf k}}$ is the momentum operator for the relative motion
${\bf r}={\bf r}_{2}-{\bf r}_{1}$, $\hat{\boldsymbol{\sigma}}_{i}$
is the spin operator of the $i$th atom, and $V\left(r\right)$ is
the interatomic potential with a finite range $\epsilon$. Our theory
may also be generalized to the case of ${\bf K}\neq0$ and ${\bf J}\neq0$,
and then more partial waves should be involved. 

One of the most daunting challenges for establishing the contact theory
is how to obtain the \emph{universal} two-body behavior at short distance
for a SO-coupled Fermi gas. Although the SO-coupled two-body problem
was considered recently by using a spherical square-well potential
\cite{Cui2012M,Wu2013S,Cui2017M}, the general form of such universal
behavior for any interatomic potential still remains elusive till
now. In this work, we develop a perturbation method to construct the
short-range asymptotic form of the two-body wave function for a SO-coupled
system. We find that two new scattering parameters $u,v$ need to
be introduced in the short-range behavior of two-body wave functions,
besides the traditional scattering length (volume) and effective ranges.
The obtained universal behavior does not depend on the short-range
details of the interatomic potentials, and thus is feasible for any
interatomic potential with short range. Two new adiabatic energy relations
are accordingly found with respect to the new scattering parameters,
i.e., 
\begin{eqnarray}
\frac{\partial E}{\partial u} & = & \frac{\hbar^{2}\lambda}{32\pi^{2}M}\left(\mathcal{C}_{a}^{(0)}-\lambda\frac{\mathcal{P}_{\lambda}}{2}\right),\label{eq:intr2}\\
\frac{\partial E}{\partial v} & = & \lambda\frac{3\hbar^{2}\mathcal{C}_{a}^{(1)}}{32\pi^{2}M},\label{eq:intr3}
\end{eqnarray}
 in which we hold all the other two-body parameters unchanged in the
partial derivatives. Here, $\mathcal{C}_{a}^{(0)}$ is the well known
$s$-wave contact, $\mathcal{C}_{a}^{(1)}$ is the $p$-wave contact
corresponding to the $p$-wave scattering volume \cite{Yu2015U,Peng2016L}.
In addition, $\mathcal{P}_{\lambda}$ is the new contact introduced
by SO coupling. Further, we derive the asymptotic behavior of the
large-momentum distribution from the universal two-body behavior at
short distance, 
\begin{multline}
n\left(q\right)=\frac{\mathcal{C}_{a}^{(1)}}{q^{2}}+\left(\mathcal{C}_{a}^{(0)}+\mathcal{C}_{b}^{(1)}+\lambda\mathcal{P}_{\lambda}\right)\frac{1}{q^{4}}+\mathcal{O}\left(q^{-6}\right),\label{eq:intr4}
\end{multline}
 in which $\mathcal{C}_{b}^{(1)}$ is the $p$-wave contact corresponding
to the $p$-wave effective range. We find that the subleading tail
($q^{-4}$) of the large-momentum distribution is amended by the new
contact $\mathcal{P}_{\lambda}$ because of SO coupling.

\emph{Universal short-range behavior of two-body wave functions}.\textemdash Let
us consider the two-body problem of a SO-coupled system in the subspace
of ${\bf K}=0$ and ${\bf J}=0$, and the corresponding Hamiltonian
takes the form of Eq.(\ref{eq:2bHam}). The subspace is spanned by
two orthogonal basis, i.e., $\Omega_{0}\left(\hat{{\bf r}}\right)=Y_{00}\left(\hat{{\bf r}}\right)\left|S\right\rangle $
and $\Omega_{1}\left(\hat{{\bf r}}\right)=-i\left[Y_{1-1}\left(\hat{{\bf r}}\right)\left|\uparrow\uparrow\right\rangle +Y_{11}\left(\hat{{\bf r}}\right)\left|\downarrow\downarrow\right\rangle -Y_{10}\left(\hat{{\bf r}}\right)\left|T\right\rangle \right]/\sqrt{3}$,
where $Y_{lm}\left(\hat{{\bf r}}\right)$ is the spherical harmonics,
$\hat{{\bf r}}$ denotes the angular part of the coordinate ${\bf r}$,
and $\left|S\right\rangle =\frac{\left(\left|\uparrow\downarrow\right\rangle -\left|\downarrow\uparrow\right\rangle \right)}{\sqrt{2}}$
and $\left\{ \left|\uparrow\uparrow\right\rangle ,\left|\downarrow\downarrow\right\rangle ,\left|T\right\rangle =\frac{\left|\uparrow\downarrow\right\rangle +\left|\downarrow\uparrow\right\rangle }{\sqrt{2}}\right\} $
are the spin-singlet and spin-triplet states with total spin $0$
and $1$, respectively. The two-body solution can formally be written
in the basis of $\left\{ \Omega_{0}\left(\hat{{\bf r}}\right),\Omega_{1}\left(\hat{{\bf r}}\right)\right\} $
as $\Psi\left({\bf r}\right)=\psi_{0}\left(r\right)\Omega_{0}\left(\hat{{\bf r}}\right)+\psi_{1}\left(r\right)\Omega_{1}\left(\hat{{\bf r}}\right)$
\cite{Cui2012M,Wu2013S}.

Since the SO effect exists even inside the interatomic potential,
it should modify the short-range behavior of the two-body wave function
dramatically \cite{Zhang2012M}. However, in current experiments of
ultracold atoms \cite{SOExperiments}, the SO-coupling strength $\lambda$
is of the order $\text{\ensuremath{\mu}m}^{-1}$, pretty small compared
to the inverse of the range of interatomic potential $\epsilon^{-1}$
(of the order nm$^{-1}$). Moreover, the momentum $k=\sqrt{ME/\hbar^{2}}$
is also much smaller than $\epsilon^{-1}$ in the low-energy scattering
limit. Therefore, when two fermions get as close as the range $\epsilon$,
we may deal with the SO coupling perturbatively as well as the energy,
and assume that the form of the two-body solution has the following
structure,
\begin{equation}
\Psi\left({\bf r}\right)\approx\phi\left({\bf r}\right)+k^{2}f\left({\bf r}\right)-\lambda g\left({\bf r}\right)\label{eq:unv1}
\end{equation}
as $r\sim\epsilon$. Here, we keep up to the first-order terms of
$k^{2}$ and $\lambda$. The advantage of this ansatz is that the
functions $\phi\left({\bf r}\right)$, $f\left({\bf r}\right)$ and
$g\left({\bf r}\right)$ are all independent on the energy and SO-coupling
strength. Therefore, they are determined only by the short-range details
of the interaction, and characterize the intrinsic properties of the
interatomic potential. We expect that the traditional scattering length
or volume in the absence of SO coupling are included in the zero-order
term $\phi\left({\bf r}\right)$, while the effective ranges are involved
in $f\left({\bf r}\right)$, the coefficient of the first-order term
of $k^{2}$. Interestingly, new scattering parameters should appear
in the first-order term of $\lambda$ (in $g\left({\bf r}\right)$),
which are introduced by SO coupling. Conveniently, more scattering
parameters may be introduced if higher-order terms of $k^{2}$ and
$\lambda$ are perturbatively considered. Inserting the ansatz (\ref{eq:unv1})
into the Schr\"{o}dinger equation, and comparing the corresponding
coefficients of $k^{2}$ and $\lambda$, we obtain 
\begin{eqnarray}
\left[-\nabla^{2}+\frac{M}{\hbar^{2}}V\left(r\right)\right]\phi\left({\bf r}\right) & = & 0,\label{eq:unv2}\\
\left[-\nabla^{2}+\frac{M}{\hbar^{2}}V\left(r\right)\right]f\left({\bf r}\right) & = & \phi\left({\bf r}\right),\label{eq:unv3}\\
\left[-\nabla^{2}+\frac{M}{\hbar^{2}}V\left(r\right)\right]g\left({\bf r}\right) & = & Q\left({\bf r}\right)\phi\left({\bf r}\right),\label{eq:unv4}
\end{eqnarray}
where $Q\left({\bf r}\right)=-i\nabla\cdot\left(\hat{\boldsymbol{\sigma}}_{2}-\hat{\boldsymbol{\sigma}}_{1}\right)$.
These equations can analytically be solved for $r>\epsilon$, and
simply yield
\begin{multline}
\Psi\left({\bf r}\right)=\alpha_{0}\left[\frac{1}{r}+\left(-\frac{1}{a_{0}}+\frac{b_{0}}{2}k^{2}+u\lambda\right)-\frac{k^{2}}{2}r\right]\Omega_{0}\left(\hat{{\bf r}}\right)\\
+\alpha_{1}\left[\frac{1}{r^{2}}+\left(\frac{k^{2}}{2}+\frac{\alpha_{0}}{\alpha_{1}}\lambda\right)+\left(-\frac{1}{3a_{1}}+\frac{b_{1}}{6}k^{2}+v\lambda\right)r\right]\Omega_{1}\left(\hat{{\bf r}}\right)\\
+\mathcal{O}\left(r^{2}\right),\label{eq:unv5}
\end{multline}
 where $\alpha_{0}$ and $\alpha_{1}$ are two complex superposition
coefficients. Apparently, $a_{0},b_{0}$ are the $s$-wave scattering
length and effective range, and $a_{1},b_{1}$ are the $p$-wave scattering
volume and effective range without SO coupling, respectively. For
simplicity, we may only consider the case with $b_{0}\approx0$ for
broad $s$-wave resonances throughout the paper. We can see that the
$s$-wave component is hybridized in the $p$-wave channel by SO coupling
as manifested as the term $\alpha_{0}\lambda/\alpha_{1}$. Interestingly,
two new scattering parameters $u$ and $v$ as we anticipate are involved.
They are the corrections from SO coupling to the short-range behavior
of the two-body wave function in $s$- and $p$-wave channels, respectively.
If $\lambda=0$, the $s$- and $p$-wave scatterings decouple, and
the asymptotic form of $\Psi\left({\bf r}\right)$ at small ${\bf r}$,
i.e., Eq.(\ref{eq:unv5}), simply reduces to the ordinary $s$- and
$p$-wave short-range boundary conditions, respectively. The derivation
above doesn't depend on the short-range details of the interaction,
and thus is universal and applicable for all kinds of neutral fermionic
atoms.

In general, the $s$- and $p$-wave scatterings in different spin
channels should both be taken into account because of SO coupling.
We may roughly estimate which partial wave is more important as follows.
Without SO coupling, and away from any resonances\textemdash in the
weak interacting limit, the two-body wave function should well behave
as ${\bf r}\rightarrow0$ as $\Psi\left({\bf r}\right)\sim\left(\alpha_{0}/a_{0}\right)\Omega_{0}\left(\hat{{\bf r}}\right)+\left(\alpha_{1}r/3a_{1}\right)\Omega_{1}\left(\hat{{\bf r}}\right)$.
If we assume that the atoms are initially prepared equally in the
spin channels $\Omega_{0}\left(\hat{{\bf r}}\right)$ and $\Omega_{1}\left(\hat{{\bf r}}\right)$,
we have $\alpha_{0}/\alpha_{1}\sim a_{0}r/3a_{1}$. When interatomic
interactions are turned on, the two-body wave function becomes divergent
as ${\bf r}\rightarrow0\left(>\epsilon\right)$, $\alpha_{0}r^{-1}$and
$\alpha_{1}r^{-2}$ for $s$- and $p$-wave scatterings, respectively.
This divergent behavior is unchanged even in the presence of SO coupling.
Then the ratio between the strengths of $s$- and $p$-wave scatterings
at small ${\bf r}$ becomes $\left(\alpha_{0}r^{-1}\right)/\left(\alpha_{1}r^{-2}\right)\approx a_{0}r^{2}/3a_{1}$.
Near $s$-wave resonances, we have $a_{0}\sim k_{f}^{-1}$, $a_{1}\sim\epsilon^{3}$,
$r\sim\epsilon$, where $k_{f}$ is the Fermi wavenumber, and then
this ratio is approximately of the order $\left(k_{f}\epsilon\right)^{-1}\gg1$.
Therefore, the $s$-wave interaction dominates the two-body scattering.
By noticing $\Omega_{0}\left(\hat{{\bf r}}\right)=\left|S\right\rangle /\sqrt{4\pi}$,
and $\Omega_{1}\left(\hat{{\bf r}}\right)=-i\left(\hat{\boldsymbol{\sigma}}_{2}-\hat{\boldsymbol{\sigma}}_{1}\right)\cdot\left({\bf r}/r\right)\left|S\right\rangle /\sqrt{16\pi}$,
and if the $p$-wave interaction could be ignored near broad $s$-wave
resonances, Eq.(\ref{eq:unv5}) becomes (up to a prefactor $\alpha_{0}/\sqrt{4\pi}$),
\begin{equation}
\Psi\left({\bf r}\right)=\left(\frac{1}{r}-\frac{1}{a_{0}}+u\lambda\right)\left|S\right\rangle -i\frac{\lambda}{2}\left(\hat{\boldsymbol{\sigma}}_{2}-\hat{\boldsymbol{\sigma}}_{1}\right)\cdot\frac{{\bf r}}{r}\left|S\right\rangle +\mathcal{O}\left(r\right),\label{eq:unv6}
\end{equation}
 which exactly recovers the result of \cite{Zhang2012M} (see Eq.(31)
of \cite{Zhang2012M}) with $a_{R}^{-1}=a_{0}^{-1}-u\lambda$. 

Near $p$-wave resonances, for example, the $p$-wave Feshbach resonance
at $B_{0}=185.09$G in $^{6}$Li \cite{Zhang2004P}, we have $a_{0}\sim\epsilon$,
$a_{1}\sim k_{f}^{-3}$, $r\sim\epsilon$, then the ratio between
the strengths of $s$- and $p$-wave scatterings is roughly of the
order $\left(k_{f}\epsilon\right)^{3}\ll1$. In this case, the $p$-wave
scattering becomes significantly important. 

\emph{Large-momentum distribution.}\textemdash For a many-body system
with $N$ spin-half fermions, if only two-body correlations are taken
into account, the many-body wave function $\Psi_{N}$ can approximately
be written as the form of Eq.(\ref{eq:unv5}), when fermions $\left(i,j\right)$
get close while all the others are far away. In this case, ${\bf r}={\bf r}_{i}-{\bf r}_{j}$,
and the arbitrary complex numbers $\alpha_{0}$ and $\alpha_{1}$
become the functions of the variable ${\bf X}$, which involves both
the center-of-mass (c.m.) coordinate of the pair being considered
and the coordinates of all the other fermions. Further, $\alpha_{0}$
and $\alpha_{1}$ should be constrained by the normalization of the
many-body wave function. Using the asymptotic form of the many-body
wave function $\Psi_{N}$ at small ${\bf r}$, we can easily obtain
the behavior of the tail of the single-particle momentum distribution
at large ${\bf q}$ (but smaller than $\epsilon^{-1}$), which is
defined as $n\left({\bf q}\right)\equiv\sum_{i=1}^{N}\int\prod_{j\neq i}d{\bf r}_{j}\left|\int d{\bf r}_{i}\Psi_{N}e^{-i{\bf q}\cdot{\bf r}_{i}}\right|^{2}$.

After straightforward algebra, we easily obtain the momentum distribution
$n\left(q\right)$ taking the form of Eq.(\ref{eq:intr4}) at large
$q\left(<\epsilon^{-1}\right)$. Here, we are only interested in the
dependence of the momentum distribution on the amplitude of ${\bf q}$,
and have already integrated over the angular part of ${\bf q}$. We
find that
\begin{equation}
\mathcal{C}_{a}^{(\nu)}=32\pi^{2}\mathcal{N}\int d{\bf X}\left|\alpha_{\nu}\left({\bf X}\right)\right|^{2},\,\left(\nu=0,1\right),\label{eq:lmd1}
\end{equation}
\begin{equation}
\mathcal{C}_{b}^{(1)}=\frac{64\pi^{2}M\mathcal{N}}{\hbar^{2}}\int d{\bf X}\alpha_{1}^{*}\left({\bf X}\right)\left[E-\hat{T}\left({\bf X}\right)\right]\alpha_{1}\left({\bf X}\right)\label{eq:lmd2}
\end{equation}
 are the conventional $s$- and $p$-wave contacts \cite{Peng2016L},
where $\hat{T}\left({\bf X}\right)$ denotes the operators of the
c.m. motion of the pair $\left(i,j\right)$ and all the other fermions,
and $\mathcal{N}=N\left(N-1\right)/2$ is the number of all the possible
ways to pair atoms. Besides, a new contacts $\mathcal{P}_{\lambda}$
resulted from SO coupling appears, which is defined as
\begin{equation}
\mathcal{P}_{\lambda}\equiv64\pi^{2}\mathcal{N}\int d{\bf X}\alpha_{0}^{*}\left({\bf X}\right)\alpha_{1}\left({\bf X}\right)+\text{c.c.}\,.\label{eq:lmd3}
\end{equation}
 Obviously, this new contact describes the interplay of the $s$-
and $p$-wave scatterings because of SO coupling.

Since the momentum distribution at large ${\bf q}$ is only characterized
by the short-range behavior of the two-body physics, we may roughly
estimate the order of all the quantities in the large-${\bf q}$ behavior
of the momentum distribution simply according to the two-body picture
as before. Near $s$-wave resonances, if initially without SO coupling
and away from any resonances, the atoms are prepared equally in the
spin states $\Omega_{0}\left(\hat{{\bf r}}\right)$ and $\Omega_{1}\left(\hat{{\bf r}}\right)$,
we have $\alpha_{0}/\alpha_{1}\sim a_{0}r/3a_{1}$, and then $\mathcal{C}_{a}^{(1)}q^{-2}/\mathcal{C}_{a}^{(0)}q^{-4}\approx9a_{1}^{2}q^{2}/a_{0}^{2}r^{2}$,
which is roughly of the order $\left(k_{f}\epsilon\right)^{4}\ll1$.
Besides, we may also find $\mathcal{C}_{b}^{(1)}/\mathcal{C}_{a}^{(0)}\sim\left(k_{f}\epsilon\right)^{4}\ll1$.
This means that the $p$-wave contribution to the tail of momentum
distribution at large ${\bf q}$ may reasonably be ignored, which
is consistent with the discussion before. However, the SO-coupling
correction is notable compared to the $p$-wave contact in the subleading
tail of the momentum distribution, i.e., $\lambda\mathcal{P}_{\lambda}/\mathcal{C}_{b}^{(1)}\sim\left(k_{f}\epsilon\right)^{-2}\gg1$. 

Near $p$-wave resonances, the leading tail $q^{-2}$ of the large-momentum
distribution becomes important, because $\mathcal{C}_{a}^{(1)}q^{-2}/\mathcal{C}_{a}^{(0)}q^{-4}\sim\left(k_{f}\epsilon\right)^{-4}\gg1$.
In the subleading tail of $q^{-4}$, we find $\mathcal{C}_{b}^{(1)}/\mathcal{C}_{a}^{(0)}\sim\left(k_{f}\epsilon\right)^{-4}\gg1$,
thus the $s$-wave contribution may be ignored. Consequently, the
momentum distribution at large ${\bf q}$ behaves as $\mathcal{C}_{a}^{(1)}q^{-2}+\left(\mathcal{C}_{b}^{(1)}+\lambda\mathcal{P}_{\lambda}\right)q^{-4}$
with a considerable correction of $\lambda\mathcal{P}_{\lambda}$
in the subleading tail due to SO coupling, compared to the $s$-wave
contribution, i.e. $\lambda\mathcal{P}_{\lambda}/\mathcal{C}_{a}^{(0)}\sim\left(k_{f}\epsilon\right)^{-2}\gg1$.

\emph{Adiabatic energy relations.}\textemdash The thermodynamics of
many-body systems, which is seemingly uncorrelated to the momentum
distribution, is also characterized by the contacts defined above.
A set of adiabatic energy relations describe how the energy of a many-body
system changes as the two-body interaction is adiabatically adjusted.
Let us consider two many-body wave functions $\Psi_{N}$ and $\Psi_{N}^{\prime}$
corresponding to different interatomic interaction strengths. From
the Schr\"{o}dinger equations satisfied by $\Psi_{N}$ and $\Psi_{N}^{\prime}$,
we easily obtain
\begin{multline}
\left(E-E^{\prime}\right)\iiint_{\mathcal{D}_{\epsilon}}d{\bf r}_{1}d{\bf r}_{2}\cdots d{\bf r}_{N}\Psi_{N}^{\prime*}\Psi_{N}=\\
-\frac{\hbar^{2}\mathcal{N}}{M}\varoiint_{r=\epsilon}{\bf I}\cdot\hat{{\bf n}}d\Sigma+\frac{\hbar^{2}\lambda\mathcal{N}}{2\pi M}\varoiint_{r=\epsilon}{\bf F}\cdot\hat{{\bf n}}d\Sigma,\label{eq:adbt1}
\end{multline}
 where ${\bf I}\equiv\Psi^{\prime*}\nabla\Psi-\left(\nabla\Psi^{\prime*}\right)\Psi$,
${\bf F}\equiv\left(\psi_{1}^{\prime*}\psi_{0}-\psi_{0}^{\prime*}\psi_{1}\right)\hat{{\bf e}}_{r}$
with the unit radial vector $\hat{{\bf e}}_{r}$ of ${\bf r}$, and
$\Psi_{N}\left({\bf X},{\bf r}\right)=\psi_{0}\left({\bf X},r\right)\Omega_{0}\left(\hat{{\bf r}}\right)+\psi_{1}\left({\bf X},r\right)\Omega_{1}\left(\hat{{\bf r}}\right)$.
Here, the domain $\mathcal{D}_{\epsilon}$ is the set of all configurations
$\left({\bf r}_{i},{\bf r}_{j}\right)$ with $r=\left|{\bf r}_{i}-{\bf r}_{j}\right|>\epsilon$,
$\Sigma$ is the surface in which the distance between the two atoms
in the pair $\left(i,j\right)$ is $\epsilon$, and $\hat{{\bf n}}$
is the direction normal to $\Sigma$ but is opposite to the radial
direction. Using the asymptotic form of the many-body wave function
$\Psi_{N}$ at small ${\bf r}$, we find
\begin{multline}
\delta E=-\frac{\hbar^{2}}{32\pi^{2}M}\left[\left(\mathcal{C}_{a}^{(0)}-\lambda\frac{\mathcal{P}_{\lambda}}{2}\right)\delta a_{0}^{-1}+\mathcal{C}_{a}^{(1)}\delta a_{1}^{-1}\right.\\
\left.-\frac{\mathcal{C}_{b}^{(1)}}{4}\delta b_{1}-\lambda\left(\mathcal{C}_{a}^{(0)}-\lambda\frac{\mathcal{P}_{\lambda}}{2}\right)\delta u-3\lambda\mathcal{C}_{a}^{(1)}\delta v\right],\label{eq:adbt2}
\end{multline}
which characterizes how the energy of the system varies as the scattering
parameters adiabatically change. In the absence of SO coupling, Eq.(\ref{eq:adbt2})
simply reduces to the ordinary form of the adiabatic energy relations
for $s$- and $p$-wave interactions \cite{Peng2016L,SWaveContacts},
with respect to the scattering length (or volume) as well as effective
range. Because of SO coupling, two new scattering parameters come
into the problem, and then additional new adiabatic energy relations
appear, i.e., Eqs.(\ref{eq:intr2})-(\ref{eq:intr3}). These adiabatic
energy relations demonstrate how the macroscopic thermodynamics of
SO-coupled many-body systems varies with microscopic two-body scattering
parameters.

\emph{Contacts} \emph{in a two-body problem}.\textemdash On behalf
of the future experiments and calculations, we may explicitly evaluate
the contacts defined above for a two-body bound state, the wave function
of which may be written as a column vector in the basis of $\left\{ \Omega_{0}\left(\hat{{\bf r}}\right),\Omega_{1}\left(\hat{{\bf r}}\right)\right\} $
as \cite{Wu2013S}
\begin{equation}
\Psi_{b}\left({\bf r}\right)=B\kappa_{-}\left[\begin{array}{c}
h_{0}^{(1)}\left(\kappa_{-}r\right)\\
-h_{1}^{(1)}\left(\kappa_{-}r\right)
\end{array}\right]+D\kappa_{+}\left[\begin{array}{c}
h_{0}^{(1)}\left(\kappa_{+}r\right)\\
h_{1}^{(1)}\left(\kappa_{+}r\right)
\end{array}\right],\label{eq:2bc1}
\end{equation}
 where $\kappa_{\pm}=i\kappa\pm\lambda$, and $\kappa=\sqrt{-ME/\hbar^{2}}$.
The binding energy $E$ can be determined by expanding $\Psi_{b}\left({\bf r}\right)$
at small ${\bf r}$ and comparing with the short-range boundary condition
(\ref{eq:unv5}), then the two-body contacts are easily obtained according
to the adiabatic relations. Near $s$-wave resonances, we find 
\begin{equation}
E=-\frac{\hbar^{2}}{Ma_{0}^{2}}+\frac{2\hbar^{2}u}{Ma_{0}}\lambda+O\left(\lambda^{2}\right),\label{eq:2bc2}
\end{equation}
 which simply reduces to the result $E=-\hbar^{2}/Ma_{0}^{2}$ in
the absence of SO coupling. Then we immediately obtain $\mathcal{C}_{a}^{(0)}=64\pi^{2}/a_{0}$
and $\mathcal{P}_{\lambda}=128\pi^{2}u$ by using adiabatic relations.
Near $p$-wave resonances, we find
\begin{equation}
E=\frac{2\hbar^{2}}{Ma_{1}b_{1}}-\frac{6\hbar^{2}v}{Mb_{1}}\lambda+O\left(\lambda^{2}\right),\label{eq:2bc3}
\end{equation}
 which is consistent with that without SO coupling \cite{Yu2015U},
and then it yields $\mathcal{C}_{a}^{(1)}=-64\pi^{2}/b_{1}$ and $\mathcal{C}_{b}^{(1)}=-256\pi^{2}\left(a_{1}^{-1}-3v\lambda\right)/b_{1}^{2}$.

\emph{Grand canonical potential and pressure relation}.\textemdash The
adiabatic relations as well as the large-momentum distribution we
obtained above is valid for any pure energy eigenstate. Therefore,
they should still hold for any incoherent mixed state statistically
at finite temperature. Then the energy, particle number density and
contacts become their statistical average values. It should be interesting
to discuss how the results presented above affect the finite-temperature
thermodynamics. To this end, let us look at the grand canonical potential,
which is defined as $\mathcal{J}\equiv-PV$ \cite{Landau2007S}, where
$P$ is the pressure and $V$ is the volume of the system. According
to straightforward dimensional analysis \cite{Braaten2008,Barth2011T},
we can obtain 
\begin{multline}
\mathcal{J}=-\frac{2}{3}E-\frac{\hbar^{2}}{96\pi^{2}Ma_{0}}\left(\mathcal{C}_{a}^{(0)}-\lambda\frac{\mathcal{P}_{\lambda}}{2}\right)\\
-\frac{\hbar^{2}\mathcal{C}_{a}^{(1)}}{32\pi^{2}Ma_{1}}+\frac{\hbar^{2}b_{1}\mathcal{C}_{b}^{(1)}}{384\pi^{2}M}+\lambda\frac{\hbar^{2}v\mathcal{C}_{a}^{(1)}}{16\pi^{2}M},\label{eq:gcp1}
\end{multline}
which alternatively yields the pressure relation by dividing both
sides of Eq.(\ref{eq:gcp1}) by $-V$.

\emph{Conclusions.}\textemdash We systematically study the contact
theory for spin-orbit-coupled Fermi gases. The universal two-body
behavior at short distance is analytically derived, by introducing
a perturbation method, which doesn't depend on the short-range details
of interatomic potentials. For simplicity, we focus on the $s$- and
$p$-wave scatterings in the subspace of vanishing center-of-mass
momentum and total angular momentum. Interestingly, two new microscopic
scattering parameters appear in the short-range behavior of two-body
wave functions because of spin-orbit coupling. We claim that this
is a general and unique feature for spin-orbit-coupled systems, and
thus the obtained universal short-range behavior of two-body wave
functions is feasible for all kinds of neutral fermionic atoms. Consequently,
a new contact is introduced originated from spin-orbit coupling, which,
combining with conventional $s$- and $p$-wave contacts, characterizes
the universal properties of spin-orbit-coupled many-body systems.
In general, more partial-wave scatterings should be taken into account
for nonzero center-of-mass momentum and nonzero total angular momentum.
Then more contacts should appear. Our method could conveniently be
generalized to other kinds of spin-orbit couplings as well as to low
dimensions. Besides, our method could also be applied to bosons. In
the presence of spin-orbit coupling, we expect that additional contacts
would be introduced for bosonic systems.
\begin{acknowledgments}
S. G. P and K. J are supported by the NKRDP (National Key Research
and Development Program) under Grant No. 2016YFA0301503 and NSFC under
Grant No. 11474315, 11674358, 11434015, 91336106. S. T is supported
by the US National Science Foundation CAREE award Grant No. PHY-1352208. 

S. G. P and C. X. Z contributed equally to this work.
\end{acknowledgments}

\end{document}